\documentclass[fleqn,10pt]{wlscirep}
\usepackage[utf8]{inputenc}
\usepackage[T1]{fontenc}
\usepackage{listings}
\usepackage{here, bm,amsmath,amssymb,cases,braket}
\usepackage{authblk}
\usepackage{ulem, color}
\title{Benchmark of quantum-inspired heuristic solvers for quadratic unconstrained binary optimization}

\author[1]{* Hiroki Oshiyama}
 
\author[1,2,3]{Masayuki Ohzeki}
\affil[1]{%
	Graduate School of Information Sciences, Tohoku University, Sendai 980-8579, Japan
}%
\affil[2]{%
	Institute of Innovative Research, Tokyo Institute of Technology, Yokohama 226-8503, Japan
}%
\affil[3]{%
	Sigma-i Co., Ltd., Tokyo 108-0075, Japan
}%

\date{\today}

\begin{abstract}
Recently, inspired by quantum annealing, many solvers specialized for unconstrained binary quadratic programming problems have been developed. For further improvement and application of these solvers, it is important to clarify the differences in their performance for various types of problems.
In this study, the performance of four quadratic unconstrained binary optimization problem solvers, namely D-Wave Hybrid Solver Service (HSS), Toshiba Simulated Bifurcation Machine (SBM), Fujitsu Digital Annealer (DA), and simulated annealing on a personal computer, was benchmarked. The problems used for benchmarking were instances of real problems in MQLib, instances of the SAT-UNSAT phase transition point of random not-all-equal 3-SAT (NAE 3-SAT), and the Ising spin glass Sherrington-Kirkpatrick (SK) model. Concerning MQLib instances, the HSS performance ranked first; for NAE 3-SAT, DA performance ranked first; and regarding the SK model, SBM performance ranked first. These results may help understand the strengths and weaknesses of these solvers.
\end{abstract}

\begin{document}

\maketitle

\section{Introduction}

Quantum annealing (QA) \cite{Farhi472, Das2008}, which is a quantum {heuristic} algorithm for solving combinatorial optimization problems, has attracted a great deal of attention because it is implemented using real quantum systems by D-Wave Systems Inc.\cite{Johnson2011, Harris2010}, aiming at becoming more powerful than classical algorithms such as simulated annealing (SA) \cite{Kirkpatrick671, Fu_1986}. %mezard1987, Hartmann2005 
To use the current D-Wave's QA device, a combinatorial optimization problem must be mapped to a quadratic unconstrained binary optimization (QUBO) problem. QUBO is an optimization problem of binary variables $x_i\in\{0,1\}$, where $i\in\{1,2,\dots,N\}$, and its cost function to be minimized is defined as
\begin{align}
E(\bm x)=\sum_{i,j}Q_{i,j}x_i x_j,
\label{eq:qubo}
\end{align}
where $Q_{i,j}$ is a real number called QUBO matrix element.
In general, QUBO is NP-hard \cite{Barahona_1982}, and many NP-complete problems and combinatorial optimization problems are mapped to QUBO \cite{Lucas2014}.

Although current QA devices have limited capability owing to hardware implementation limitations, in anticipation of future developments of QA devices, methods using QUBO models for solving real-world problems in a variety of fields have been actively studied \cite{Perdomo-Ortiz2012, PhysRevLett.108.230506, Babbush_2014, venturelli2016, Mott2017,PhysRevX.7.041052, Li2018}.
Inspired by this trend, several sophisticated heuristic QUBO solvers have been developed and commercialized \cite{HSS_WP,Inagaki603,Gotoeaav2372, Aramon_2019}. It is highly non-trivial to determine whether a particular algorithm is more powerful than another because the performance of heuristic algorithms varies depending on the target problem. For successful application to real-world problems and further development of these QUBO solvers, it is necessary to clarify the strengths and weaknesses of each solver for various types of QUBO problems. In this study, we benchmarked the performance of three commercialized QUBO solvers including one using a real QA device: D-Wave Hybrid Solver Service (HSS), Toshiba Simulated Bifurcation Machine (SBM), and Fujitsu Digital Annealer (DA). {In order to understand the characteristics of the solvers, we benchmark various types of problems, including Ising spin glass problems and real-world problems. This is in contrast to a similar benchmark study reported recently\cite{kowalsky20213regular}, which used only a single kind of constraint satisfaction problem (specifically, 3-regular 3-XORSAT). While in Ref. \cite{kowalsky20213regular}, the size dependence of the time to obtain an optimal solution with a certain probability is analyzed in detail, in this study, the performance of the solvers is evaluated by comparing the value of the cost function obtained for a given execution time. Such a performance evaluation will be helpful in application cases where approximate solutions are acceptable.}

The remainder of this paper is organized as follows. In Sec. \ref{sec:solver}, we briefly explain the solvers benchmarked. In Sec. \ref{sec:prob}, the definition of the problem instances used for benchmarking are provided. In Sec. \ref{sec:result}, we present the results of the benchmarking experiment. {Concluding remarks are given in Sec. \ref{sec:sum}.}

\section{QUBO solvers} \label{sec:solver}
In this section, we briefly explain the four solvers used in this study. Three commercial solvers were benchmarked. For comparison, we also experimented with SA on a personal computer. 

The first solver is HSS, commercialized by D-Wave Systems Inc.\cite{HSS_WP}. This solver is a so-called quantum-classical hybrid algorithm that employs QA as an accelerator. Note that the actual implementation of the algorithm is not open to the public.
Thus, it is unclear how QA is used internally. We used HSS hybrid BQM solver, version 2.0, which can manage up to $10^6$ variables and $2\times10^8$ couplings \cite{HSS_WP2}. {We accessed HSS via Leap cloud.}

The second solver is SBM, commercialized by Toshiba \cite{Gotoeaav2372}. {The QA inspired algorithm of SBM, so-called simulated bifurcation (SB) algorithm,} uses the adiabatic time evolution of Kerr-nonlinear parametric oscillators (KPOs) \cite{goto2016}. The dynamics in the classical limit of KPOs can be quickly computed in classical computers by solving the independent equations of motion in parallel \cite{Gotoeaav2372}. { To overcome accuracy degradation caused by analog errors due to the use of the dynamics of continuous variables, a variant of the SB algorithm called ballistic SB (bSB) algorithm was developed, which mitigate the analog error by modifying the potential term of the equation of motion. As a further improvement of the bSB algorithm, the discrete SB (dSB) algorithm was also developed, which reduces the analog error by discretizing the potential term of the bSB algorithm \cite{goto2021}.} We use SBM evaluation version {1.5.1} (which is not publicly available), that {uses dSB algorithm and} can manage all-to-all coupling of up to {$10^6$} variables and $10^8$ nonzero couplings. Parallelization is 80 or 160 per GPU. In this study, we used the \texttt{autoising} solver; hyperparameters are automatically searched by the solver. We accessed SBM via evaluation version directly from Toshiba.

The third solver is DA, commercialized by Fujitsu \cite{Aramon_2019}. DA uses an SA-specific hardware architecture to accelerate {the parallel tempering Markov chain Monte Carlo (MCMC) calculation} \cite{Matsubara2018, Tsukamoto2017}. Although DA does not use quantum algorithms, it is inspired by D-Wave devices in the sense that the hardware is specialized for QUBO solving. We used \texttt{fujitsuDA2PT} solver, which can manage all-to-all coupling of up to 8192 variables. {We accessed DA via DA Center Japan}.

For comparison with these commercial solvers, we ran SA using the open-source software D-Wave neal, version 0.5.7 \cite{github_neal}, on a personal computer {with Ubuntu 20.04.3 LTS and Python 3.8.2}. {D-Wave neal neal implements SA with MCMC without parallel tempering method}. The CPU used in the experiment was Intel Core i9-9900K, and single-threaded runs were performed.

%SBM, DA 2021 1-2
%HSS 2020 9/19-20 NAE 2020 12 SK 2021 2/25

%\subsection{D-Wave Hybrid Solver Service}
%\subsection{Toshiba Simulated Bifurcation Machine}
%\subsection{Fujitsu Digital Annealer}
%\subsection{Simulated annealing}

\section{Problem instances for benchmarking} \label{sec:prob}
In this section, we explain the three problem sets used in the conducted benchmarking.

\subsection{MQLib repository instances}
 We used the same set of 45 problems used in the benchmarks presented in HHS’s white paper \cite{HSS_WP,github_hss}. This problem set is extracted from the MQLib repository, and some of the problems have their origin in real-world problems, such as image segmentation  \cite{DunningEtAl2018}. This problem set was reported to be time-consuming to solve because of all the heuristics contained in the MQLib library. Concerning benchmarking, a 20-minute run is recommended for each problem. \cite{HSS_WP}. 
 The 45 problems are uniformly classified into nine classes: three classes according to size (small: $1000\leq N\leq 2500$, medium: $2500< N\leq 5000$, and large: $5000< N\leq 10000$) and three classes according to edge density (sparse: $d\leq 0.1$, medium: $0.1<d\leq0.5$, and dense: $0.5<d$), where $d$ is the number of edges divided by the number of edges in a complete graph of the same size \cite{HSS_WP}.
 
\subsection{Not-All-Equal 3-SAT}
{Satisfiability problem (SAT) is one of the most fundamental NP-hard problems and therefore it is good benchmark problem for heuristic solvers.} Not-all-equal 3-SAT (NAE 3-SAT) is a variant of the Boolean SAT problem and is an NP-complete problem \cite{darmann2019}. NAE 3-SAT requires at least one literal to be true and at least one literal to be false in each clause with three literals. The cost function of a random NAE 3-SAT with $N$ variables and $M$ clauses is expressed in a straightforward manner in the Ising model with $\sigma_i \in\{-1,1\}$, where $1\leq i\leq N$:
\begin{align}
    E(\bm \sigma)=\frac{1}{4}\sum_{m=1}^{M}(\zeta_{{m,1}}\zeta_{{m,2}}\sigma_{i_{m,1}}\sigma_{i_{m,2}}+\zeta_{{m,2}}\zeta_{{m,3}}\sigma_{i_{m,2}}\sigma_{i_{m,3}}+\zeta_{{m,3}}\zeta_{{m,1}}\sigma_{i_{m,3}}\sigma_{i_{m,1}}+1), \label{eq:nae3sat}
\end{align}
where $i_{m,l}\in\{1,2,\dots,N\}$ and $\zeta_{m,l}\in\{-1,1\}$ for $1\leq m\leq M$ and $1\leq l\leq 3$ are random variables that follow a discrete uniform distribution; $\zeta_{m,l}=-1$ corresponds to the negation of the $l$-th Boolean variable in clause $m$. Each clause has three different variables, i.e., $i_{m,l}\neq i_{m,l^{\prime}}$ if $l\neq l^{\prime}$. If the minimum of $E(\bm\sigma)$ in Eq. (\ref{eq:nae3sat}) is 0 for a given formula, it is satisfiable (SAT); otherwise, it is unsatisfiable (UNSAT). The QUBO formulation as in Eq. (\ref{eq:qubo}) can be easily obtained from this Ising formulation by the variable transformation $x_i=(\sigma_i+1)/2$. {Because NAE 3-SAT has such a natural QUBO representation, it is a suitable benchmark problem for QUBO solvers amongst SAT variants.} When the clause-to-variable ratio is $M/N= 2.11$, the SAT-UNSAT phase transition occurs, and problem instances are most difficult to solve \cite{Achlipotas2001, 10.1007/3-540-61551-2_70, Gent1994TheSP}. In this study, we used randomly generated instances with this critical clause-to-variable ratio for benchmarking.

\subsection{Sherrington-Kirkpatrick model}
The Sherrington-Kirkpatrick (SK) model is an Ising spin glass model with infinite spatial dimensions \cite{Sherrington1975,mezard1987}. The cost function of $N$ variables with no external field is expressed as
\begin{align}
    E(\bm\sigma)=\frac{1}{\sqrt N}\sum_{1\leq i<j\leq N}J_{i,j}\sigma_i\sigma_j,
\end{align}
where $J_{i,j}$ is a random Gaussian variable. As previously explained, the QUBO formulation can be easily obtained. The mean field analysis shows that the energy landscape of the SK model has a many-valley structure separated by asymptotically infinitely large energy barriers, which implies that it is extremely difficult to find the exact solution \cite{tap1977}. In this study, we used randomly generated instances with $J_{i,j}$ presenting zero mean and unity standard deviation for benchmarking.

\section{Results}\label{sec:result} % {and discussions}
In this section, we present benchmarking results for each of the three problem sets introduced in the previous section. In the results shown below, the network time required to send the instance and receive the result was ignored in the measurement of execution time. Regarding HSS, the number of seconds specified in \texttt{time\_limit} was used as the execution time. For SBM, the time specified in \texttt{timeout} was used as the execution time. Concerning DA, there was no parameter to specify the execution time directly, so \texttt{total\_elapsed\_time} recorded in the response file was used as the execution time. Finally, for SA with D-Wave neal, we measured the time taken for the \texttt{sample} function to finish.

\subsection{MQLib instances}
First, we present the results for a 5-min experiment of the instances from the MQLib repository. For HSS and SBM, the execution time was set to 5 min. For DA, \texttt{number\_replicas} was set to $128$ and \texttt{number\_iterations} was adjusted for each instance so that the deviation of execution time in 5 min was within 20 s. 
Concerning SA, \texttt{num\_sweeps} was adjusted for each instance such that the execution time was 5 min. 

\begin{figure}[t]
\begin{center}
\includegraphics[width=0.9\textwidth]{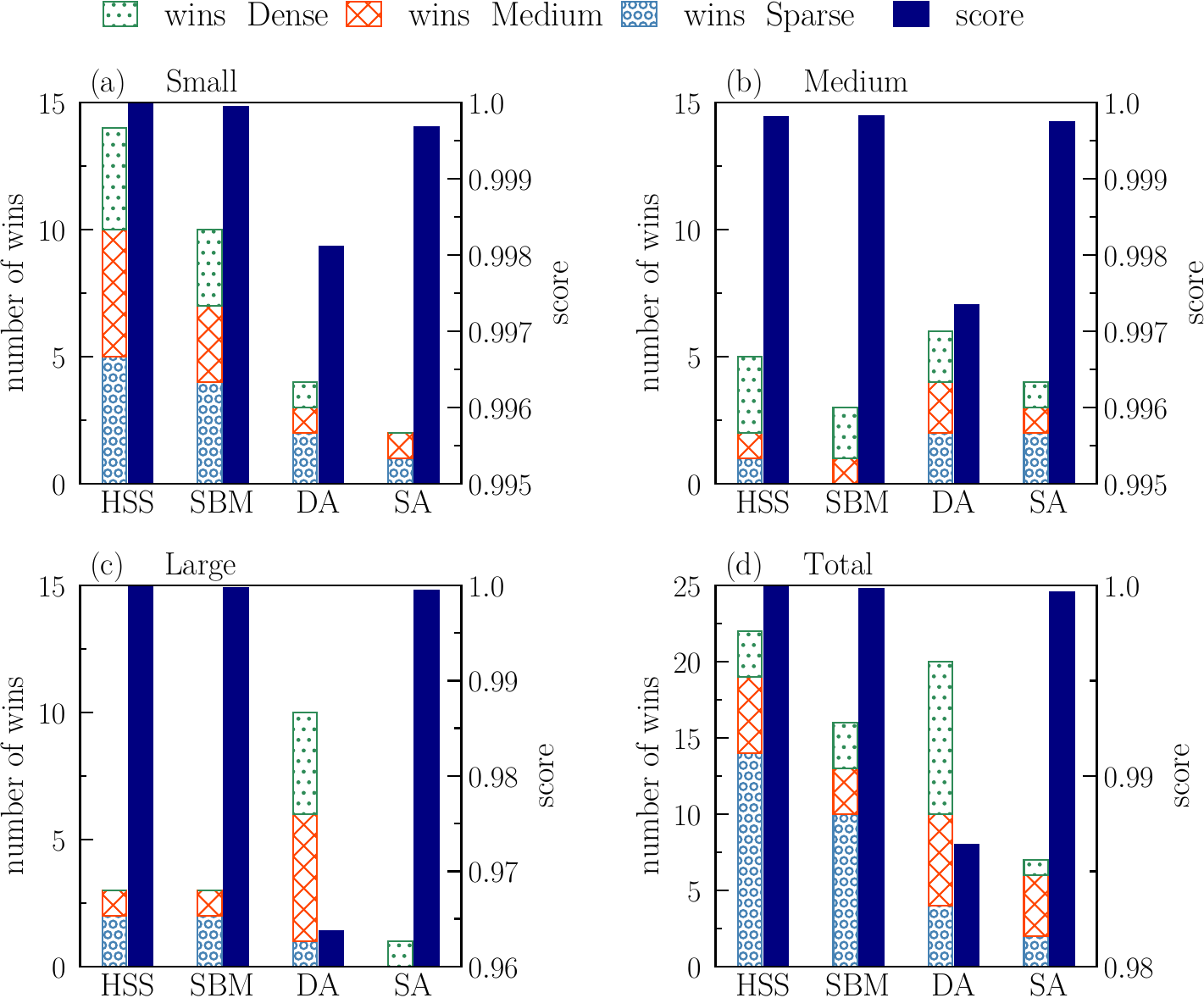}
\end{center}
\caption{{Number of wins (left axis) and average score (right axis) for a 5-min experiment of MQLib instances. Each panel shows the result for a class categorized by problem size, (a) Small, (b) Mediun, and (c) Large; (d) Total number of wins and average score for all instances. The score for each instance is defined by Eq. (\ref{eq:score}). In calculating the average score, instance g000644 was ignored due to absence of data for DA.}
\label{fig:mqlib}}
\end{figure}

%Table \ref{tabel:win}
{Figure \ref{fig:mqlib}} shows the number of wins for each solver; this number was counted when the solver obtained the best solution. If there was more than one solver with the best solution, the number of wins was counted for all of them. The total result for all classes was that HHS won most of the problems (22), followed by DA (20), SBM (16), and SA (7). The results for each class classified by size show that HSS won the most for the small class, while DA won the most for the medium and large classes. The results for each class classified by edge density show that, for Sparse class, HSS won the most, for Medium class, DA won the most, and HSS and DA won the most. The number of wins of SA was only 2 at most, and most of the time, it was 0 or 1 for each of the nine classes. 

{Furthermore, we evaluate the quality of the obtained solution using a score defined as the ratio of the value of cost function ($E_{\mathrm{solver}}=\{E_\mathrm{HSS},~E_\mathrm{SBM},~E_\mathrm{DA}~,E_\mathrm{SA} \}$) to the best value obtained in this experiment ($E_0=\min\{E_\mathrm{HSS},~E_\mathrm{SBM},~E_\mathrm{DA}~,E_\mathrm{SA} \}$):
\begin{align}
    S_{\mathrm{solver}}=E_{\mathrm{solver}}/E_0~~(\mathrm{solver}\in\{\mathrm{HSS},~\mathrm{SBM},~\mathrm{DA}~,\mathrm{SA} \})\label{eq:score}.
\end{align}
Tables \ref{tabel:cost0}-\ref{tabel:cost8} show the score for each instance, and Fig. \ref{fig:mqlib} shows the average of the scores for Small, Medium, and Large classes, and for all instances. The original lowest values of the cost function found in this benchmarking are listed in Table \ref{tabel:cost_full}. The average scores of HSS and SBM are almost identical and higher than other solvers. This implies that HSS and SBM have stable performance on a wide range of problems. On the other hand, DA has an exceptionally bad solution for the instance g001345, which is why the average score drops significantly in the Large class. In addition, in the Small and Medium classes, the average score of DA is about 0.01 lower than the other solvers. This implies that DA is slightly less stable, becuase even for SA, which has the fewest wins, the difference in average score from HSS is within 0.001.} 
\newpage
%Tables \ref{tabel:cost0}-\ref{tabel:cost8} show the value of the cost function obtained by each solver for each instance; this value is the ratio of the best value found in this experiment. For almost all instances, the solutions by HSS and SBM were in agreement by more than three orders of magnitude with the best solution found in this experiment. The solutions by DA tended to have larger differences from the best solution compared to other solvers. The lowest values of the cost function found in this benchmarking are listed in Table \ref{tabel:cost_full}.

%\begin{table}[t]
%\begin{center}
%    \input{mqlib_5min_wins}
%  \caption{Number of wins for a 5-min experiment of MQLib instances. The row direction is classified in terms of size (Small, Medium, and Large), whereas the column direction is classified in terms of edge density (Sparse, Medium, and Dense). \label{tabel:win}}
%\end{center}
%\end{table}

\begin{table}[H]
\begin{center}
\begin{tabular}[t]{lll|llll}\hline
    input & size & density & HSS & SBM & DA & SA\\\hline\hline
g000989 & 2319 & 0.00086 & $\bm{1.0}$ & $\bm{1.0}$ & $\bm{1.0}$ & 0.998708\\\hline
g003215 & 2206 & 0.00093 & $\bm{1.0}$ & 0.999457 & 0.998103 & 0.997985\\\hline
g001269 & 2294 & 0.0017 & $\bm{1.0}$ & $\bm{1.0}$ & $\bm{1.0}$ & 0.999847\\\hline
g000421 & 2034 & 0.0038 & $\bm{1.0}$ & $\bm{1.0}$ & 0.985010 & 0.999303\\\hline
g002440 & 2242 & 0.044 & $\bm{1.0}$ & $\bm{1.0}$ & 0.999213 & $\bm{1.0}$\\\hline

\end{tabular} %Values of cost functions, ratio to the best solution found in a 5-min experiment of MQLib instances
  \caption{Values of score, defined by Eq. (\ref{eq:score}) for Small and Sparse classes. The first row shows the instance name, the second row presents the number of variables, the third row contains the edge density, and the fourth and subsequent rows show the results for each solver. The values are computed in single precision from the obtained solution of binary variables; they are shown with six decimal places. The best solutions obtained in this benchmarking are shown in bold. \label{tabel:cost0}}
\end{center}
\end{table}

\begin{table}[H]
\begin{center}
\begin{tabular}[t]{lll|llll}\hline
    input & size & density & HSS & SBM & DA & SA\\\hline\hline
g000432 & 2153 & 0.11 & $\bm{1.0}$ & 0.999958 & 0.999045 & 0.999974\\\hline
g000524 & 2218 & 0.14 & $\bm{1.0}$ & $\bm{1.0}$ & $\bm{1.0}$ & $\bm{1.0}$\\\hline
g002586 & 2079 & 0.16 & $\bm{1.0}$ & $\bm{1.0}$ & 0.999102 & 0.999890\\\hline
g001327 & 2318 & 0.3 & $\bm{1.0}$ & $\bm{1.0}$ & 0.999300 & 0.999928\\\hline
g001469 & 2412 & 0.46 & $\bm{1.0}$ & 0.999824 & 0.998105 & 0.999911\\\hline

\end{tabular}
  \caption{Results for Small and Medium classes, same as Table \ref{tabel:cost0}\label{tabel:cost1}.}
\end{center}
\end{table}

\begin{table}[H]
\begin{center}
\begin{tabular}[t]{lll|llll}\hline
    input & size & density & HSS & SBM & DA & SA\\\hline\hline
g002600 & 2432 & 0.85 & $\bm{1.0}$ & 0.999999 & 0.999505 & 0.999976\\\hline
g000969 & 2453 & 0.86 & $\bm{1.0}$ & $\bm{1.0}$ & 0.995138 & 0.999645\\\hline
g002898 & 2041 & 0.86 & $\bm{1.0}$ & $\bm{1.0}$ & $\bm{1.0}$ & 0.999996\\\hline
g001581 & 2383 & 0.86 & 0.999999 & $\bm{1.0}$ & 0.999640 & 1.000000\\\hline
g000788 & 2342 & 0.88 & $\bm{1.0}$ & 0.999860 & 0.999492 & 0.999959\\\hline

\end{tabular}
  \caption{Results for Small and Dense classes, same as Table \ref{tabel:cost0}\label{tabel:cost2}.}
\end{center}
\end{table}

\begin{table}[H]
\begin{center}
\begin{tabular}[t]{lll|llll}\hline
    input & size & density & HSS & SBM & DA & SA\\\hline\hline
g000377 & 3398 & 0.00069 & 0.998763 & 0.999890 & $\bm{1.0}$ & 0.998353\\\hline
g002569 & 2815 & 0.0011 & $\bm{1.0}$ & 0.999459 & 0.983877 & 0.998564\\\hline
g001086 & 3706 & 0.0016 & 0.998913 & 0.998673 & 0.985686 & $\bm{1.0}$\\\hline
g001337 & 2850 & 0.051 & 0.999975 & 0.999923 & $\bm{1.0}$ & 0.999931\\\hline
g000283 & 3364 & 0.072 & 0.999946 & 0.999905 & 0.997073 & $\bm{1.0}$\\\hline

\end{tabular}
  \caption{Results for Medium and Sparse classes, same as Table \ref{tabel:cost0}\label{tabel:cost3}.}
\end{center}
\end{table}

\begin{table}[H]
\begin{center}
\begin{tabular}[t]{lll|llll}\hline
    input & size & density & HSS & SBM & DA & SA\\\hline\hline
g002512 & 4731 & 0.12 & 0.999913 & 0.999861 & $\bm{1.0}$ & 0.999980\\\hline
g000802 & 3956 & 0.13 & 0.999990 & $\bm{1.0}$ & 0.998449 & 0.999919\\\hline
g003059 & 3447 & 0.14 & 0.999973 & 0.999939 & $\bm{1.0}$ & 0.999962\\\hline
g002332 & 3181 & 0.22 & 0.999994 & 0.999996 & 0.999156 & $\bm{1.0}$\\\hline
g002034 & 2528 & 0.35 & $\bm{1.0}$ & 0.999997 & 0.999201 & 0.999979\\\hline

\end{tabular}
  \caption{Results for Medium and Medium classes, same as Table \ref{tabel:cost0}\label{tabel:cost4}.}
\end{center}
\end{table}

\begin{table}[H]
\begin{center}
\begin{tabular}[t]{lll|llll}\hline
    input & size & density & HSS & SBM & DA & SA\\\hline\hline
g003198 & 3972 & 0.74 & $\bm{1.0}$ & 0.999956 & 0.999616 & 0.999979\\\hline
g002207 & 2677 & 0.74 & $\bm{1.0}$ & $\bm{1.0}$ & $\bm{1.0}$ & 0.999954\\\hline
g001913 & 3865 & 0.75 & $\bm{1.0}$ & 0.999786 & 0.999333 & 0.999643\\\hline
g001393 & 3938 & 0.83 & 0.999967 & $\bm{1.0}$ & $\bm{1.0}$ & 0.999886\\\hline
g002370 & 3884 & 0.84 & 0.999716 & 0.999843 & 0.997744 & $\bm{1.0}$\\\hline

\end{tabular}
  \caption{Results for Medium and Dense classes, same as Table \ref{tabel:cost0}\label{tabel:cost5}}
\end{center}
\end{table}

\begin{table}[H]
\begin{center}
\begin{tabular}[t]{lll|llll}\hline
    input & size & density & HSS & SBM & DA & SA\\\hline\hline
imgseg-216041 & 7724 & 0.00039 & $\bm{1.0}$ & 0.999919 & 0.996163 & 0.995890\\\hline
imgseg-376020 & 7455 & 0.00049 & $\bm{1.0}$ & 0.999522 & 0.989190 & 0.998811\\\hline
g001883 & 6831 & 0.00059 & 1.000000 & $\bm{1.0}$ & 0.999489 & 0.999998\\\hline
g000644 & 10000 & 0.0016 & 0.999307 & $\bm{1.0}$ &  & 0.999752\\\hline
g000476 & 8000 & 0.002 & 0.999457 & 0.999766 & $\bm{1.0}$ & 0.999860\\\hline

\end{tabular}
  \caption{Results for Large and Sparse classes, same as Table \ref{tabel:cost0}\label{tabel:cost6}. For input g001883, HSS and SBM had almost the same value of the cost function, while the solution configurations were truly different from each other. The result of DA for input g000644 is blank because DA can only manage 8192 variables.}
\end{center}
\end{table}

\begin{table}[H]
\begin{center}
\begin{tabular}[t]{lll|llll}\hline
    input & size & density & HSS & SBM & DA & SA\\\hline\hline
g002312 & 6395 & 0.19 & 0.999957 & 0.999054 & $\bm{1.0}$ & 0.999930\\\hline
g002563 & 6279 & 0.19 & 0.999842 & 0.999966 & $\bm{1.0}$ & 0.999945\\\hline
g000495 & 5438 & 0.21 & 0.999941 & 0.999980 & $\bm{1.0}$ & 0.999958\\\hline
g002204 & 5368 & 0.44 & $\bm{1.0}$ & $\bm{1.0}$ & $\bm{1.0}$ & 0.999903\\\hline
g000503 & 5046 & 0.45 & 0.999954 & 0.999966 & $\bm{1.0}$ & 0.999983\\\hline

\end{tabular}
  \caption{Results for Large and Medium classes, same as Table \ref{tabel:cost0}.\label{tabel:cost7}}
\end{center}
\end{table}

\begin{table}[H]
\begin{center}
\begin{tabular}[t]{lll|llll}\hline
    input & size & density & HSS & SBM & DA & SA\\\hline\hline
g002527 & 5378 & 0.59 & 0.999949 & 0.999574 & $\bm{1.0}$ & 0.999885\\\hline
g001345 & 5066 & 0.74 & 0.999252 & 0.999004 & 0.475147 & $\bm{1.0}$\\\hline
p7000-2 & 7001 & 0.8 & 0.999992 & 0.999748 & $\bm{1.0}$ & 0.999563\\\hline
g002300 & 5038 & 0.94 & 0.999970 & 0.999988 & $\bm{1.0}$ & 0.999995\\\hline
g001651 & 5819 & 0.97 & 0.999949 & 0.999913 & $\bm{1.0}$ & 0.999930\\\hline

\end{tabular}
  \caption{Results for Large and Dense classes, same as Table \ref{tabel:cost0}.\label{tabel:cost8}}
\end{center}
\end{table}

\subsection{NAE 3-SAT instances}
Next, we present the results for the random NAE 3-SAT instances with a number of variables $N=8192$ and a number of clauses $M=17285$, i.e., instances with a clause-to-variable ratio $N/M\approx2.11$. Figure \ref{fig:nae3sat} (a) shows the average of ten randomly generated instances of the cost function as a function of the execution time. As a reference, Fig. \ref{fig:nae3sat} (b) shows the results for ten different instances. Given that each data point was obtained from an independent run, a longer run may lead to a worse solution than a shorter run. In the range 100-600 s, DA presented the lowest value of the cost function, closely followed by SBM and SA; HSS presented the highest value.
{In the region below 100 s, SBM and SA showed lower energy than DA. After a long-time calculation of about 1000 seconds, HSS finally reaches the same performance as SBM and SA, but still not as good as the result of 100 second run of DA. Interestingly, the performance of SBM and {SA} is almost identical for a wide range of execution time.
}
%Focusing on the slope in the region beyond 200 s, HSS was the steepest, while SBM and SA were almost flat.
%Therefore, it is likely that HSS can obtain a much better solution than SBM and SA in a longer calculation. DA presented a steeper slope than SBM and SA in the region that we could calculate. In summary, SBM and SA are better for short-time calculations, while DA and HSS are better for long-time calculations. The lowest values of the cost function in this experiment were obtained using DA.

\subsection{SK model}
\begin{figure}[t]
\begin{center}
\begin{tabular}{cc}
\includegraphics[width=.48\textwidth]{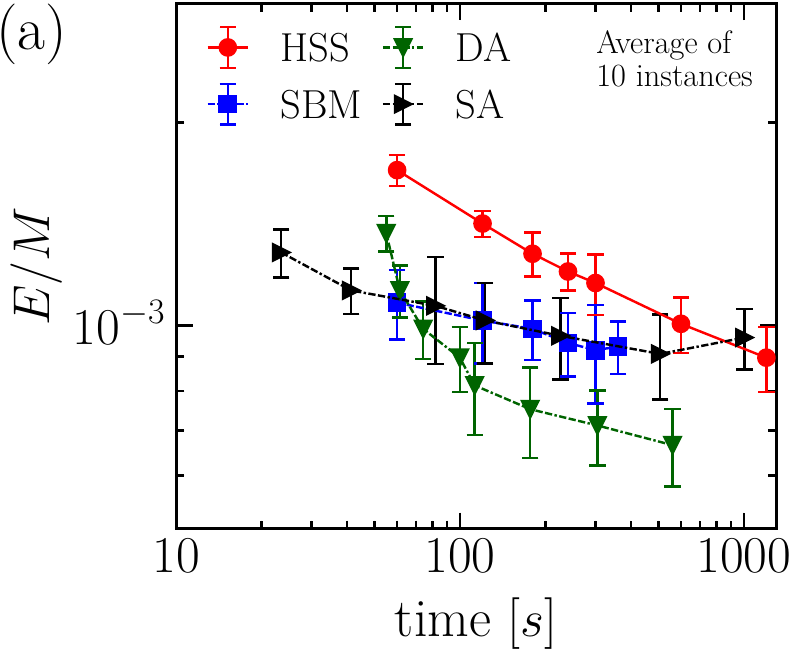}
&
\includegraphics[width=.48\textwidth]{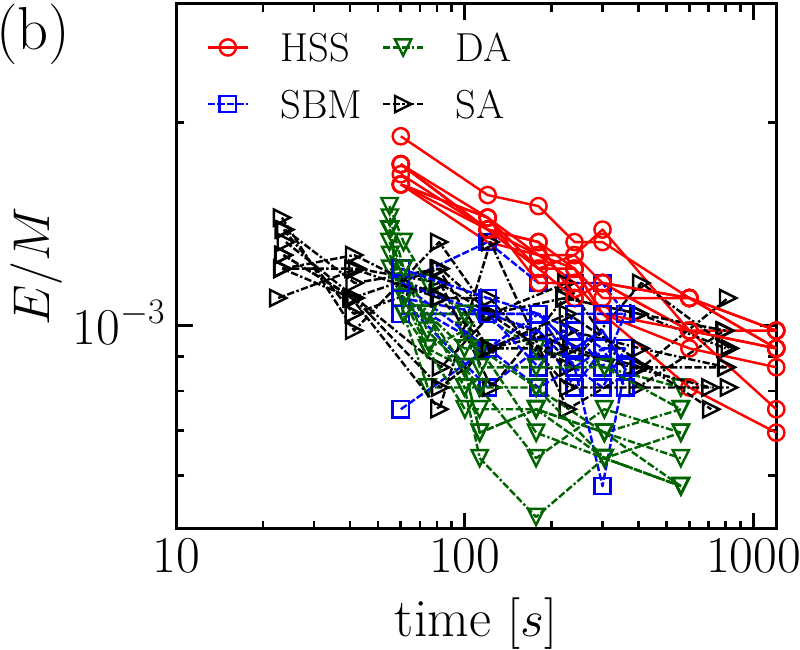}
\end{tabular}
\end{center}
\caption{
Value of the cost function per clause as a function of the execution time, obtained for NAE 3-SAT with a number of variables $N=8192$ and a number of clauses $M=17285$, i.e., $M/N\approx2.11$. Each data point was obtained from an independent run. See the main text for the time metric of each solver. (a) Average of ten instances. The error bars denote standard deviation. For DA and SA, the execution time was also averaged. (b) Results for ten different instances. 
\label{fig:nae3sat}}

\begin{center}
\begin{tabular}{cc}
\includegraphics[width=.48\textwidth]{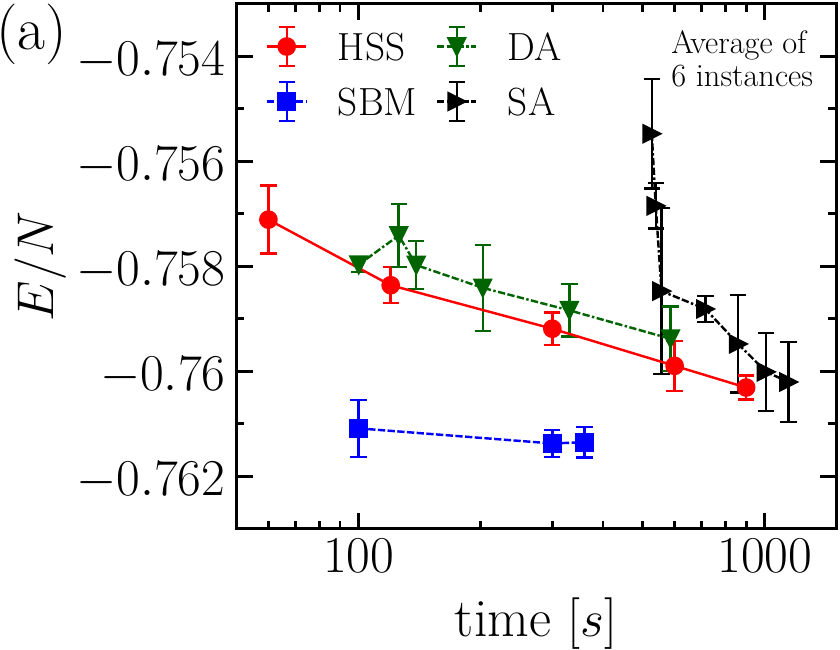}
&
\includegraphics[width=.48\textwidth]{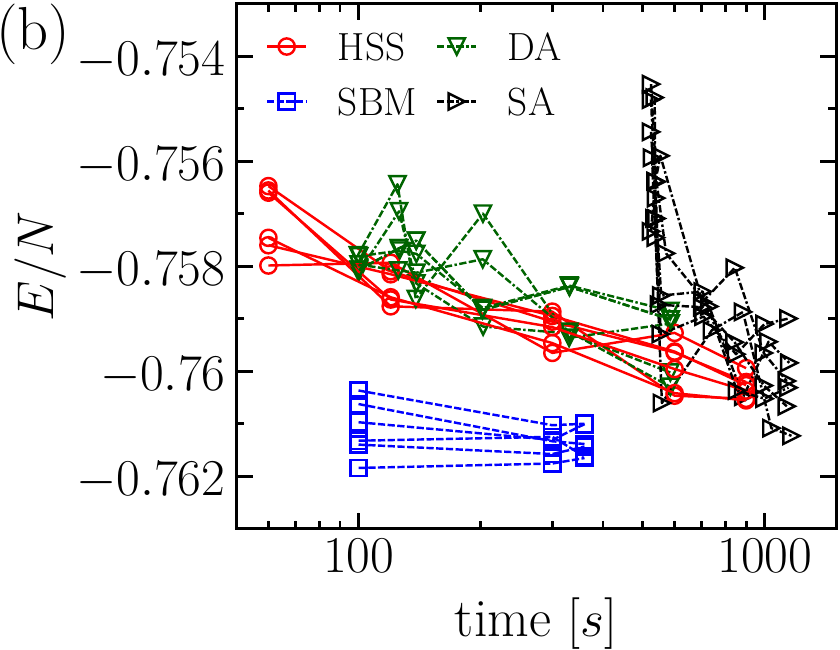}
\end{tabular}
\end{center}
\caption{
Value of cost function per variable as a function of the execution time, obtained for the SK model with a number of variables $N=8192$ and $J=1$. Each data point was obtained from an independent run. See the main text for the time metric of each solver. (a) Average of ten instances. The error bars denote standard deviation. For DA and SA, the execution time was also averaged. (b) Results for six different instances. 
\label{fig:SK}}
\end{figure}

Finally, we present the results for the SK model with 8192 variables. As with the NAE 3-SAT instances, the experiments were performed by varying the execution time. Figure \ref{fig:SK} (a) shows the average of six randomly generated instances of the cost function as a function of the execution time. As a reference, Fig. \ref{fig:SK} (b) shows the results for six different instances. SBM {clearly outperformed the other solvers}, achieving the best solutions at the 100-s mark, with little energy change for longer runs. HSS and DA showed almost the same time dependence, although HSS provided a slightly better solution. {It is interesting that this pair is different from the pair, SBM and SA, that exhibits similar performance in NAE 3-SAT instances.} In runs longer than 600 s, SA obtained as good solutions as HSS and DA, but due to the all-to-all coupling, its pre-processing calculation was expensive, requiring at least approximately 500 s for the total calculation time. %In summary, SBM performs the best in terms of both computation time and solution quality. {Just like SBM and SA exhibit similar performance in NAE 3-SAT, the performance of HSS and DA is similar over a wide range of execution times.} %HSS and DA show similar performance .

\begin{table}[b]
\begin{center}
\begin{tabular}[t]{lll}\hline
    input & value of cost function & solvers\\\hline\hline
g000989  & -2322 & HSS, SBM, DA\\\hline
g003215  & -821734 & HSS\\\hline
g001269  & -45661 & HSS, SBM, DA\\\hline
g000421  & -41680.2 & HSS, SBM\\\hline
g002440  & -2000460 & HSS, SBM, SA\\\hline
g000432  & -188363.1 & HSS\\\hline
g000524  & -4335188 & HSS, SBM, DA, SA\\\hline
g002586  & -7161694 & HSS, SBM\\\hline
g001327  & -9267492 & HSS, SBM\\\hline
g001469  & -1.42273e+07 & HSS\\\hline
g002600  & -41194.45 & HSS\\\hline
g000969  & -6647406 & HSS, SBM\\\hline
g002898  & -1.276648e+07 & HSS, SBM, DA\\\hline
g001581  & -730413.1 & SBM\\\hline
g000788  & -1962898 & HSS\\\hline
g000377  & -445529 & DA\\\hline
g002569  & -5.084731e+08 & HSS\\\hline
g001086  & -3819.935 & SA\\\hline
g001337  & -4634430 & DA\\\hline
g000283  & -337340.8 & SA\\\hline
g002512  & -327679.6 & DA\\\hline
g000802  & -2819460 & SBM\\\hline
g003059  & -3782885 & DA\\\hline
g002332  & -4586683 & SA\\\hline
g002034  & -698788.1 & HSS\\\hline
g003198  & -1.373565e+08 & HSS\\\hline
g002207  & -6781175 & HSS, SBM, DA\\\hline
g001913  & -1177002 & HSS\\\hline
g001393  & -358732 & SBM, DA\\\hline
g002370  & -5.622634e+07 & SA\\\hline
imgseg-216041  & -9572357 & HSS\\\hline
imgseg-376020  & -1.376284e+07 & HSS\\\hline
g001883  & -403013.1 & SBM\\\hline
g000644  & -132820 & SBM\\\hline
g000476  & -106794 & DA\\\hline
g002312  & -2.867864e+07 & DA\\\hline
g002563  & -5.848182e+07 & DA\\\hline
g000495  & -1.638467e+07 & DA\\\hline
g002204  & -1.229112e+08 & HSS, SBM, DA\\\hline
g000503  & -8.506962e+07 & DA\\\hline
g002527  & -8261389 & DA\\\hline
g001345  & -4.011876e+07 & SA\\\hline
p7000-2  & -1.824995e+07 & DA\\\hline
g002300  & -9.409027e+07 & DA\\\hline
g001651  & -130005.8 & DA\\\hline

\end{tabular}
  \caption{The lowest values of cost function found in this benchmarking for MQLib instances.\label{tabel:cost_full}}
\end{center}
\end{table}

\section{{{Discussion and} Conclusion}}\label{sec:sum}
We benchmarked the {heuristic} QUBO solvers, HSS, SBM, DA, and SA, using the instances from the MQLib repository, random NAE 3-SAT, and the SK model. Benchmarking with problems of various origins revealed some of the characteristics of the strengths and weaknesses of each solver. 
{For MQLib instances, which are a set of various problem instances including real-world problems, HSS showed the best performance on average, and SBM also showed stable performance that was not so different from HSS. DA outperformed other solvers on large instances, but it gave slightly poor solutions to some instances. It is rather natural result that the performance of DA varied depending on the instances because the performance of heuristic algorithms strongly depends on the problem instances in general, and it is somewhat surprising that HSS and SBM showed stable performance.
In this experiment, with a run time of 5 minutes, we find that the difference in the value of cost function of the obtained solutions is often less than 0.01\%, which is probably negligible in some application cases. Therefore, a possible direction for further study is to investigate how the results change in experiments with shorter run times.
For random NAE 3-SAT instances at the SAT-UNSAT transition point, which is a typical hard optimization problem, DA performed best for most of the execution times. The performance of SBM and SA was almost the same, and HSS was the worst. It is believed that local search methods such as the parallel tempering method used in DA do not work well for SAT instances at the SAT-UNSAT transition point that have very few solutions\cite{10.1007/3-540-61551-2_70}, and there is probably no efficient algorithm. Therefore, the result that DA still performed best implies that other solvers are also not particularly effective, which is as expected.
For SK model, which is a typical hard problem originated from the spin glass, SBM exhibited a clear advantage over other solvers, while HSS and DA showed similar performance. Since the parallel tempering method is considered to work relatively well for the SK model, it is a bit surprising that SBM, rather than DA, showed outstanding performance as opposed to the case of NAE 3-SAT. } 
% PT \cite{Machta2009}
It is an important challenge to understand the characteristics of each solver found in this study from the viewpoint of their algorithm and hardware architecture.

\section*{Acknowledgements}
We thank Murray Thom, Catherine McGeoch, Hayato Goto, and Yoshihiko Nishikawa for fruitful discussion on our benchmark tests.
In addition, we acknowledge research supports on various aspects from D-Wave Systems Inc. and TOSHIBA CORPORATION.
M. O. thanks financial support from JSPS KAKENHI Grant Number 20H02168, the Next Generation High-Performance Computing Infrastructures and Applications R $\&$ D Program by MEXT, and MEXT-Quantum Leap Flagship Program Grant Number JPMXS0120352009.

\section*{{Data availability}}
{All other data used in this study are available from the corresponding authors upon reasonable request. The problem instaces of MQlib is available from the MQLib repository \cite{github_mqlib}. 
The NAE 3-SAT and SK model instance was generated reproduced by the python program shown in Listings \ref{list:nae3sat} and \ref{list:sk} with Python 3.8.2 on Ubuntu20.04.3 LTS.
}

\begin{lstlisting}[language=Python, caption=Python program that generate the NAE 3-SAT instances used in this study in Ising formulation. Ten seed values from 0 to 9 were used. ,  label=list:nae3sat]
import numpy
import random
import itertools
seed = 0 # in range(10)
random.seed(a=seed)
N = 8192  # number of variables
M = 17285 # number of clauses
variables = range(0, N)
signs = range(-1, 2, 2)
J = {} # Ising interaction
for i in range(M):
    clause = random.sample(variables, 3)
    negations = random.choices(signs, k=3)
    v_pairs = itertools.combinations(clause, 2)
    s_pairs = itertools.combinations(negations, 2)
    for pair, sign in zip(v_pairs, s_pairs):
        J[pair] = J.get(pair, 0) + numpy.prod(sign)
\end{lstlisting}

\begin{lstlisting}[language=Python, caption=Python program that generate the SK model instances used in this study in Ising formulation. Six seed values from 1 to 6 were used. , label=list:sk]
import numpy
import random
seed = 0 # in range(6)
random.seed(a=seed)
N = 8192  # number of variables
J = {} # Ising interaction
for i in range(N):
    for j in range(i+1,N):
        J[(i, j)] = random.gauss(0, 1)/numpy.sqrt(N)
\end{lstlisting}

\bibliography{ref}
%\appendix
%\section{Problem instances used in the benchmark}
\end{document}